\documentclass[twoside,10pt,letterpaper]{article}
\usepackage{amsmath}
\usepackage{graphicx}

\newcommand{\bea}{\begin{eqnarray}}

\newcommand{\eea}{\end{eqnarray}}

\begin{document}
%

\title{The unreasonable effectiveness of equilibrium-like
theory for interpreting non-equilibrium experiments}
\author{R. Dean Astumian\\ Department of Physics and Astronomy\\ University of
Maine\\ Orono, Maine 04469-5709\\ e-mail:
astumian@maine.edu\vspace{1ex}l}
\date{}

\maketitle


\pagestyle{myheadings}


\begin{abstract}
There has been great interest in applying the results of statistical mechanics to single molecule experiements.  Recent work has highlighted so-called non-equilibrium work-energy relations and Fluctuation Theorems which take on an equilibrium-like (time independent) form.  Here I give a very simple heuristic example where an equilibrium result (the barometric law for colloidal particles) arises from theory describing the {\em thermodynamically} non-equilibrium phenomenon of a single colloidal particle falling through solution due to gravity.  This simple result arises from the fact that the particle, even while falling, is in {\em mechanical} equilibrium (gravitational force equal the viscous drag force) at every instant. The results are generalized by appeal to the central limit theorem.  The resulting time independent equations that hold for thermodynamically non-equilibrium (and even non-stationary) processes offer great possibilities for rapid determination of thermodynamic parameters from single molecule experiments.
\end{abstract}
 \vspace{5mm}
\maketitle

\noindent The equilibrium distribution for colloidal particles in dilute aqueous suspension follows a  familiar barometric or exponential law  \cite{perrin_jacp,ito_lang}

\begin{equation}
c(h) = c(0) \, \exp{(\frac{-m \, g \, h}{k_B \, T})}
\end{equation}

\noindent where $c(h)$ is the concentration of particles at height $h$.  For spherical particles of radius $r$ the effective mass is $ m = 4 \pi  r^3 (\rho_p - \rho_w)/3$ where $\rho_p$ and $\rho_w$ are the mass densities of the particle and of water, respectively.  {\em Eq. (1) is an equilibrium result that implicitly involves many particles so that the concentrations (particle densities) are well defined. } 

We can look at the situation of colloids from the very different perspective of a single particle falling through solution.  The forces acting are gravity, $m g$, viscous drag $\gamma v \sim \eta r v$ (where $\eta \approx 10^{-3}$ kg/(m s) is the viscosity of water), and a random thermal noise force the origin of which is the molecular movement of the water molecules. If we wait sufficiently long (about $m/\gamma = 10^{-6}$ s for a micron sized particle) the particle will reach terminal velocity $v_{\rm term} = m g/\gamma$ where the force of gravity is balanced by the viscous drag and there is no further acceleration.   Then, taking as the origin of the coordinate system the center of mass of the particle at some time $t = 0$ after terminal velocity is attained, the subsequent probability density function $P(h,t)$ for the particle position can be written

\begin{equation}
P(h,t) = \frac{\exp{[\frac{-(h -(m g /\gamma) t)^2}{4 D t}]}}{\sqrt{4 \pi D t}}
\end{equation}

\noindent which is a solution of Fick's equation for diffusion with drift \cite{berg_rand}

\begin{equation}
\frac{\partial{P(h,t)}}{\partial t} = D \frac{\partial^2 P(h,t)}{\partial h^2} + \frac{m g}{\gamma} \frac{\partial P(h,t)}{\partial h}
\end{equation}
  
\noindent where D is the diffusion coefficient.  Eq. 2 describes a Gaussian distribution with mean $\mu =  (m g/\gamma) t$ and variance $\sigma^2 = \langle h(t)^2 \rangle - \langle h(t) \rangle^2 = 2 D t$.   Although it is more likely of course that the particle moves downward, at short times, there is a reasonable chance that thermal noise will cause the particle to be found slightly higher than where it started at $t=0$.  Since the distance from the center of the Gaussian to the original $h = 0$ is $[(m g/\gamma) t]$, the probability for a particle to be above its starting point at time t is $\frac{1}{2}{\rm erfc}\,[\frac{((m g /\gamma) t)^2}{4 D t}]$ where ${\rm erfc}\,(x)$ is the complement of the error function.  

These ``upward" trajectories have been termed ``violations of the second law of thermodynamics on short time and small length scales" \cite{evans_prl} because the entropy change is negative for trajectories where colloidal particles move up.  The upward trajectories {\em certainly do NOT} demonstrate a violation of the more relevant Thomson (Lord Kelvin) formulation of the second law which states that no repeatable (cyclic) process can do work on the environment with the sole change being a decrease in the temperature of the system.   The upward trajectories can however be exploited in a non-isotropic system to allow the random input of energy to drive directed motion by a Brownian motor mechanism \cite{astumian_sci97}.

We can calculate exactly the ratio of the probability density for the particle to ``fall up" to $+h$ to the probability density to fall down to $-h$ from Eq. 2 to be 

\begin{equation}
\frac{P(h,t)}{P(-h,t)} = \exp{(\frac{- m g h}{\gamma D})}
\end{equation}

\noindent Remarkably, time has disappeared altogether in this ratio, and by inserting Einstein's relation \cite{einstein} $\gamma D = k_B T$ we regain the ``equilibrium" barometric law Eg. 1!  An analogous relation for motion of an overdamped particle in an arbitrary potential was derived in a more general context by Bier et al. \cite{bier} using Onsager's \cite{onsager} thermodynamic action approach.

Since the probability density function is normalized  $$\int_{-\infty}^{+\infty}  P(h) dh =  \int_{-\infty}^{+\infty}  P(-h) dh =1$$  we also have 

\begin{equation}
\langle \exp{(\frac{ m g h}{k_B T})}\rangle = \int_{-\infty}^{+\infty}\ exp{(\frac{ m g h}{k_B T})} P(h) dh =  1
\end{equation}

\noindent The quantity $W_{\rm diss} = -mgh$ is the work dissipated when a particle falls a distance h.  

Many stochastic processes obey a Gaussian distribution due to the central limit theorem \cite{berg_pnas} which states that the probability density function for any variable that can be expressed as the sum of many small random quantities approaches a Gaussian distribution irrespective of the microscopic laws governing the individual small quantities so long as the mean and variance of the small quantities are finite.  Consequently we expect equilibrium-like (time independent) relations to hold for a wide range of processes.  

Consider an arbitrary macroscopic variable $q$, often termed a generalized displacement, and the time derivative of its average  $\dot{\overline{q}}$ which is often called a generalized velocity or flux.  If $q$ fluctuates due to molecular noise in the environment the distribution for $q$ will be given by 

\begin{equation}
P(q,t) = \frac{\exp{[\frac{-(q - \dot{\overline{q}} t)^2}{4 D t}]}}{\sqrt{4 \pi D t}}
\end{equation}

 \noindent where the variance $\sigma^2 = \langle q(t)^2 \rangle -  \langle q(t) \rangle^2 = 2 D t$ and we take $q(0) = 0$.  As before $P(q,t) = P(-q,t) \exp{[q \dot{\overline{q}}/D]}$ and 
 
 \begin{equation}
 \langle \exp{[q \dot{\overline{q}}/D]} \rangle = 1.
 \end{equation}
 
\noindent Further, if the generalized velocity $\dot{\overline{q}}$ is related linearly to a generalized force $X$ such that $\dot{\overline{q}} = X/\gamma$ (as is the case after terminal velocity is reached) we have $P(q,t) = P(-q,t) \exp{[q X/ (\gamma D)]}$.  Finally, with identification of $\gamma D = k_B T$ (a  fluctuation-dissipation relation) we have  

\begin{equation}
\frac{P(q)}{P(-q)} =  \exp{[(W_{\rm diss}/(k_B T)]}
\end{equation}

\noindent where as before the dissipated work $W_{\rm diss}$ is the product of the generalized force and displacement $-q X$.  Using the thermodynamic relation that the dissipated work is the difference between the total work and the stored free energy $W_{\rm diss} = W - \Delta G$ we have the equality

\begin{equation} 
\langle \exp{(W/(k_B T))} \rangle = \exp(\Delta G/k_B T)
\end{equation}

\noindent Relations analogous to equations (8) and (9) (termed non-equilibrium fluctuation-dissipation theorems (FDT)) derived from a different perspective were published twenty five years ago by Bochkov and Kozovlev \cite{boch_physa81}. 

Interest in the non-equilibrium FDT has recently been rekindled by the  theoretical work of Jarzynski \cite{jar_prl97}, Bier et al. \cite{bier},  Crooks \cite{crooks_pre99},  and others.  This work has made important progress by relating experimental observables from non-equilibrium experiments to thermodynamic parameters such as free energy.  Hummer and Szabo \cite{szabo} pointed out that the relations are implicit in the Feynman-Kac path integral theorem, and demonstrated how the non-equilibrium work energy relations and fluctuation theorems can be used to interpret single molecule experiments. The relations (8) and (9) have been experimentally tested by Bustamante and colleagues \cite{bust_nat,bust_pt,bust_sci}  who showed that a relatively small number of single molecule experiments can be used to obtain very good estimates of the Equilibrium Free Energy profiles for the system. 

It is often stated that many single molecule experiments are carried out far from equilibrium.  Indeed, the 10-100 nN force typical for an AFM pulling experiment is enormous compared to the, say, 10-20 pN force associated with even strong molecular motors \cite{howard_book01} and the rate of change of the force - up to several hundred nN per second - seems very large.  However, when we compare the rate of change of the force ($dF/dt \sim 100$ nN/s) with the ratio of the characteristic force \cite{purcell} $F_{\rm char} = \eta^2/\rho \sim 10^{-9}$N  to the characteristic relaxation time for a nanometer object in solution $t_{\rm char} = \rho\,\, {r}^2/\eta \propto m/\gamma \sim 10^{-12} {\rm s}$   ($F_{\rm char}/t_{\rm char} \sim 10^3$ N/s) we see that the system is in fact very close to mechanical equilibrium at every instant in time.  The characteristic quantities are written in terms of the particle size $r \sim 10^{-9} m$, density $\rho \sim 10^3 kg/m^3$, and the viscosity of the solution $\eta \sim 10^{-3} kg/(m s)$. Nevertheless, the system is far from thermodynamic equilibrium as it continues to move, and to dissipate energy in the environment.  

In nanoscale physics, it is essential to remember that ``equilibrium" does not have a single unambiguous meaning.  The  thermodynamic {\em non-equilibrium} aspect of a typical experimental system shows up as a mean energy dissipation rate that is not zero.  The mechanical {\em equilibrium} aspect shows up in the Gaussian distribution of fluctuations about the mean dissipated work (or mean extension) over many realizations.  The MECHANICAL equilibrium allows a simple description of the distribution of the fluctuations of quantities about their mean values in THERMODYNAMICALLY far from equilibrium situations.  The closeness to mechanical equilibrium is ultimately the explanation for the unreasonable effectiveness of equilibrium theory for interpreting experiments which, from our perspective (but not from that of the single molecule!) appear to be ``far from equilibrium".

Time independent equations that hold for non-equilibrium (and even non-stationary) processes offer great possibilities for rapid determination of thermodynamic parameters from single molecule experiments.   The present paper is intended to provide a simple background for understanding these equilibrium-like relations, and shine light on their historical origins to facilitate further investigation.
\\
{\em Acknowledgements}  I am very grateful to Peter Hanggi for pointing out the paper of Bochkov and Kuzovlev, and for enlightening discussions on the origins of generalized fluctuation relations; I am grateful to many colleagues, including Chris Jarzynski, Jim McClymer,Jan Liphardt, Peter von Hippel, Neil Comins, John Schellman, Howard Berg, Dudley Herschbach, John Ross, Bob Adair, Hans Frauenfelder, Michel Peyrard, Bob Mazo, Ioan Kosztin, Steve Block, Joe Howard, Ken Dill, and Hans von Baeyer, for reading and commenting on the manuscript.

\end{document}